\documentclass[sigconf, authorversion, nonacm]{acmart}

\AtBeginDocument{%
  }

\usepackage{adjustbox}
\usepackage{xspace}
\usepackage{xurl}                                 
\usepackage{graphicx}                             
\usepackage{natbib}                               
\usepackage{booktabs}                             
\usepackage{paralist}                             
\usepackage{mathtools}
\usepackage[framemethod=TikZ]{mdframed} 
\usepackage{multirow}
\usepackage{listings}
\usepackage{caption}
\usepackage{xcolor}

\usepackage[capitalize,sort&compress]{cleveref}
\usepackage{enumitem}
\usepackage[T1]{fontenc}  
\usepackage{textcomp}  
\usepackage{wrapfig}
\usepackage{subcaption}
\usepackage{rotating}
\usepackage{pifont}
\usepackage{siunitx}
\usepackage{colortbl}

\definecolor{set19c1}{HTML}{e41a1c} 
\definecolor{set19c2}{HTML}{377eb8} 
\definecolor{set19c3}{HTML}{4daf4a} 
\definecolor{set19c4}{HTML}{984ea3} 
\definecolor{set19c5}{HTML}{ff7f00} 
\definecolor{set19c6}{HTML}{ffff33} 
\definecolor{set19c7}{HTML}{a65628} 
\definecolor{set19c8}{HTML}{f781bf} 
\definecolor{set19c9}{HTML}{999999} 

\definecolor{set39c1}{HTML}{8dd3c7} 
\definecolor{set39c2}{HTML}{ffffb3} 
\definecolor{set39c3}{HTML}{bebada} 
\definecolor{set39c4}{HTML}{fb8072} 
\definecolor{set39c5}{HTML}{80b1d3} 
\definecolor{set39c6}{HTML}{fdb462} 
\definecolor{set39c7}{HTML}{b3de69} 
\definecolor{set39c8}{HTML}{fccde5} 
\definecolor{set39c9}{HTML}{d9d9d9} 

\definecolor[named]{ACMBlue}{cmyk}{1,0.1,0,0.1.}
\definecolor[named]{ACMYellow}{cmyk}{0,0.16,1,0}
\definecolor[named]{ACMOrange}{cmyk}{0,0.42,1,0.01}
\definecolor[named]{ACMRed}{cmyk}{0,0.90,0.86,0}
\definecolor[named]{ACMLightBlue}{cmyk}{0.49,0.01,0,0}
\definecolor[named]{ACMGreen}{cmyk}{0.20,0,1,0.19}
\definecolor[named]{ACMPurple}{cmyk}{0.55,1,0,0.15}
\definecolor[named]{ACMDarkBlue}{cmyk}{1,0.58,0,0.21}

\definecolor{snsblue}{HTML}{4C72B0}
\definecolor{snsorange}{HTML}{DD8452}
\definecolor{snsgreen}{HTML}{55A868}
\definecolor{snsred}{HTML}{C44E52}
\definecolor{snspurple}{HTML}{8172B3}
\definecolor{snsbrown}{HTML}{937860}
\definecolor{snspink}{HTML}{DA8BC3}
\definecolor{snsgrey}{HTML}{8C8C8C}
\definecolor{snsgold}{HTML}{CCB974}
\definecolor{snscyan}{HTML}{64B5CD}

\colorlet{tableblue}{set39c5!50}
\colorlet{tablegreen}{set39c7!50}

\colorlet{hlsnsblue}{snsblue!25}
\colorlet{hlsnsorange}{snsorange!25}
\colorlet{hlsnsgreen}{snsgreen!25}
\colorlet{hlsnsred}{snsred!25}
\colorlet{hlsnspurple}{snspurple!25}
\colorlet{hlsnsbrown}{snsbrown!25}
\colorlet{hlsnspink}{snspink!25}
\colorlet{hlsnsgrey}{snsgrey!25}
\colorlet{hlsnsgold}{snsgold!25}
\colorlet{hlsnscyan}{snscyan!25}

\DeclareSIUnit{\exec}{execs}
\DeclareSIUnit{\month}{mth}
\DeclareSIUnit{\year}{yr}
\DeclareSIUnit{\cpuhour}{CPU\textrm{-}hr}
\DeclareSIUnit{\cpuyear}{CPU\textrm{-}yr}
\DeclareSIUnit{\loc}{LoC}

\lstdefinestyle{x86assembly}{
  language=[x86masm]Assembler,
  morekeywords={mov, movq, movsxd, xor, shr, inc, adc},
  deletekeywords={eax, edx, al},
}

\newlist{inlineroman}{enumerate*}{1}
\setlist[inlineroman]{label=(\roman*)}

\newlist{inlinealph}{enumerate*}{1}
\setlist[inlinealph]{label=(\alph*)}

\newlist{researchq}{enumerate}{1}
\setlist[researchq]{
  label=\textbf{RQ~\arabic*},
  ref=\arabic*,
}
\crefname{researchqi}{RQ}{RQs}
\Crefname{researchqi}{RQ}{RQs}

\newcommand{\aflpp}{AFL++\xspace}
\newcommand{\pear}{\textsc{PeAR}\xspace}
\newcommand{\gtirbrewriting}{\texttt{gtirb-rewriting}\xspace}

\newcommand{\cmark}{{\color{set19c3}\ding{51}}\xspace}
\newcommand{\xmark}{{\color{set19c1}\ding{55}}\xspace}
\newcommand{\afldyninst}{\texttt{afl-dyninst}\xspace}
\newcommand{\ddisasm}{\texttt{Ddisasm}\xspace}
\newcommand{\fuzzbench}{\textsc{FuzzBench}\xspace}
\newcommand{\icicle}{\textsc{Icicle}\xspace}
\newcommand{\retrowrite}{\texttt{RetroWrite}\xspace}
\newcommand{\stochfuzz}{\textsc{StochFuzz}\xspace}
\newcommand{\zafl}{\textsc{ZAFL}\xspace}

\newcommand{\shaderow}{\rowcolor{lightgray!10}}

\newcommand{\totalfuzztime}{4.25~CPU-yrs\xspace}
\newcommand{\numevalsbi}{four\xspace}

\newenvironment{hlbox}[1]{ 
  \mdfsetup{ 
    linecolor=snsgrey, 
    linewidth=2pt, 
    roundcorner=2pt, 
    frametitle={\colorbox{hlsnsgrey}{\space\textbf{#1}\space}}, 
    frametitleaboveskip=\dimexpr-\ht\strutbox\relax, 
    innertopmargin=0pt, 
    skipabove=0.5\baselineskip, 
    skipbelow=0.5\baselineskip, 
    leftmargin=5pt, 
    rightmargin=5pt, 
    nobreak=true, 
  } 
  \begin{mdframed}%
}{ 
  \end{mdframed} 
}

\begin{document}

\title{\pear: A Static Binary Rewriting Framework for Binary-Only Fuzzing}

\author{Alvin Charles}
\affiliation{
\institution{School of Computing\\The Australian National University}
\city{Canberra}
\state{ACT}
\country{Australia}
}

\author{Adrian Herrera}
\affiliation{
\institution{School of Computing\\The Australian National University}
\city{Canberra}
\state{ACT}
\country{Australia}
}

\author{Peter Oslington}
\affiliation{
\institution{School of Computing\\The Australian National University}
\city{Canberra}
\state{ACT}
\country{Australia}
}

\author{Alwen Tiu}
\affiliation{
\institution{School of Computing\\The Australian National University}
\city{Canberra}
\state{ACT}
\country{Australia}
}

\begin{abstract}
Binary-only fuzzing is a key technique for finding bugs in close-source software.
Without access to source code, the fuzzer must rely on static or dynamic binary
instrumentation for coverage guidance. In practice, most fuzzers favor dynamic
binary instrumentation~(DBI), accepting runtime overhead to avoid the perceived
accuracy and soundness challenges associated with static binary
instrumentation~(SBI). We show that these concerns are unwarranted, and that
accurate, scalable~SBI is achievable using off-the-shelf frameworks. Building on
these frameworks, we develop \pear, an extensible binary-only fuzzing framework.
We demonstrate \pear's versatility by implementing several modern fuzzer
features---including, deferred initialization, persistent mode, and
shared-memory fuzzing.

We evaluate \pear over~\totalfuzztime of fuzzing on the \fuzzbench benchmark and
find that \pear:
\begin{inlineroman}
\item successfully instruments \qty{88}{\percent} of \fuzzbench targets,
comparable to the best SBI-based fuzzers;

\item achieves a median throughput improvement of~$4\times$ when using
persistent mode and shared memory fuzzing; and

\item attains coverage comparable to compiler-based instrumentation.
\end{inlineroman}
Our results show that~SBI is a practical and effective technique for binary-only
fuzzing, and that modern binary rewriting frameworks can apply complex
instrumentation with high granularity and negligible performance compromise.
\end{abstract}



\maketitle

\section{Introduction}
\label{sec:introduction}

Fuzzing is a popular and effective bug-finding technique.  It relies on the
ability to quickly generate a large number of inputs and test those inputs
against a target program.  
These inputs are generated to explore corner cases in
the target's behavior, which existing tests may not cover and where bugs may
lurk. 
Modern \emph{greybox} fuzzers rely on a feedback loop to determine whether these
corner cases are being reached.  Feedback is derived from program
instrumentation that measures and tracks code coverage, guiding the fuzzer to
uncover and explore new target behaviors.  This instrumentation is typically
injected at compile time.  However, there are often cases where source code is
unavailable (e.g., when testing close-source software), requiring alternative
techniques for tracking coverage.

\emph{Binary-only fuzzing}---fuzzing when source code is unavailable---relies
on either static binary instrumentation (SBI) or dynamic binary instrumentation
(DBI) to track code coverage.  SBI frameworks apply instrumentation to target
binaries by generating new binaries (e.g., via rewriting) containing the desired
instrumentation.  SBI has a relatively low runtime performance cost (compared to~DBI).
However, reliably and correctly applying instrumentation statically is
challenging, with many~SBI frameworks requiring strong assumptions about the
target binary~\cite{bauman2018superset}.

In contrast, DBI applies instrumentation to the target at runtime.  DBI
techniques include:
\begin{inlineroman}
\item just-in-time (JIT) recompilation (e.g., Intel Pin~\cite{luk2005pin}, Dyninst~\cite{dyninst});

\item emulation frameworks (e.g., QEMU~\cite{bellard2005qemu}, \icicle~\cite{chesser2023icicle}); and

\item probe-based instrumentation (e.g., \texttt{kprobes} in the Linux kernel).
\end{inlineroman}
While applying~DBI is simpler (compared to~SBI), it comes with significant
performance costs (e.g., $10\textrm{--}100\times$ when fuzzing the~LAVA-M
benchmark suite with~AFL~\cite{dolan2016lava}).  Thus, improving the accuracy
and generalizability of SBI has seen a renewed focus.

Despite advances in SBI fuzzing frameworks~\cite{dinesh2020retrowrite,%
  gao2021scalable,zhang2021stochfuzz,nagy2021breaking}, we found that many still
lack the features and extensibility of modern compiler- and DBI-based systems.
Key performance techniques---such as deferred initialization, persistent mode,
and shared memory fuzzing---remain unsupported. 
Moreover, we found that many of these frameworks failed to generalize and
accurately instrument common fuzz targets, despite recent
works~\cite{zhang2021stochfuzz,schulte2020gtirb} showing that accurate static
binary rewriting is now practical. This raises the question: can these advances
be used to implement advanced fuzzing techniques in binary-only fuzzers?

We answer this question by developing \pear, an efficient binary-only fuzzing
framework that implements advanced fuzzing techniques. \pear uses an off-the-shelf
static binary rewriter based on the GrammaTech Intermediate Representation for
Binaries~\cite{schulte2020gtirb} (GTIRB) to inject fuzzing instrumentation with
features and performance comparable to state-of-the-art compiler-based solutions.
In summary, we contribute:
\begin{itemize}
\item \pear, an~SBI fuzzing framework for injecting modern fuzzing instrumentation
across a wide range of targets (\cref{sec:pear});

\item A comprehensive evaluation (over \totalfuzztime) of \pear and
\numevalsbi other state-of-the-art SBI fuzzing frameworks: \\
 \zafl~\cite{nagy2021breaking}, \stochfuzz~\cite{zhang2021stochfuzz},
\afldyninst~\cite{vanheuser2021afldyninst}, and E9AFL~\cite{gao2021scalable}, 
on the \fuzzbench benchmark suite (\cref{sec:evaluation}).
\end{itemize}

Our results show that \pear achieves the highest code coverage among SBI-based
fuzzers while being the only SBI framework to support advanced fuzzing
techniques.  With persistent mode and shared memory fuzzing enabled, \pear
achieves a median throughput improvement of~$4\times$ over its base
configuration, all while achieving comparable levels of code coverage to
compiler-based instrumentation.
\pear is available as an open-source project.\footnote{https://github.com/pear-labs/PeAR.git}

\section{Background and Related Work}
\label{sec:background}

\subsection{Greybox Fuzzing}
\label{sec:background-greybox-fuzzing}

Fuzzing is an important technique for automated bug discovery.
Fuzzers typically require a target program and a set of valid inputs---known as
``seeds''---to bootstrap the fuzzer. In a fuzzing campaign, new inputs are
generated from these seeds (typically via random mutation) and are tested
against the target. If the target crashes, the crashing input is saved for
further root-cause analysis post campaign.

Coverage-guided greybox fuzzing extends this idea by instrumenting the
target to collect coverage information during program execution.  Collected
coverage information is then used to further guide input generation.  This
instrumentation must have low runtime overhead, allowing the fuzzer to maintain
a high throughput and maximizing the number of inputs the fuzzer executes during
a campaign.

In addition to low-overhead instrumentation, modern greybox fuzzers---popularized
by \aflpp~\cite{fioraldi2020afl++}---use the following techniques to boost
performance and maximize throughput.

\textbf{Forkserver.}
Greybox fuzzers must repeatedly execute the target with new inputs.  However,
spawning a new process for each target is a costly operation.  Moreover, time is
wasted waiting for the target process to initialize before an input is executed.
To avoid this overhead, modern fuzzers use a \emph{forkserver} that ensures the
target is only initialized once. Future processes are then efficiently cloned
(due to \texttt{fork}'s copy-on-write semantics) from the initialized process.

\textbf{Deferred initialization.}
Target binaries may have complex initialization procedures that slow down the
fuzzer. To combat this, the forkserver can be initialized at a user-specified
location.  This \emph{deferred initialization} can be used to skip
target-specific initialization overheads.

\textbf{Persistent mode.}
Rather than spawning a new process per input, \emph{persistent mode} allows one
process to execute multiple inputs.  This is achieved by wrapping the relevant
target code in a loop, with each loop iteration executing a single input.  While
this increases fuzzer throughput, care must be taken to correctly reset the
target's state after executing each input (preventing future inputs from executing in
an invalid state).

\textbf{Shared memory fuzzing.}
Fuzzers typically send inputs to the target via the filesystem.  The target is
then expected to open and handle this file.  Reading and writing these inputs to
the filesystem incurs a performance penalty.  However, \emph{shared memory fuzzing}
allows the target to read inputs directly from memory, avoiding interaction with
the filesystem.

\subsection{Static Binary Instrumentation}
\label{sec:sbi}

\begin{table*}
\centering
\footnotesize

\caption{Comparison of SBI fuzzing frameworks.  We compare frameworks across two
dimensions: the rewriter they use, and the fuzzing features they support.}
\label{tab:sbi-compare}

\begin{tabular}{l ll cccc}
\toprule
& 
\multicolumn{2}{c}{Rewriter} &
\multicolumn{4}{c}{Fuzzing} \\
\cmidrule(lr){2-3}
\cmidrule(lr){4-7}
\multirow{-2}{*}{Name} &
Name & 
Type &
Forkserver &
Defer.\ init. &
Pers. mode &
Sh. mem. fuzzing \\
\midrule
\shaderow \afldyninst~\cite{vanheuser2021afldyninst} & Dyninst~\cite{dyninst} & Trampoline & \cmark & \xmark & \xmark & \xmark \\
E9AFL~\cite{gao2021scalable} & E9~\cite{duck2020binary} & Trampoline & \cmark & \xmark & \xmark & \xmark \\
\shaderow \retrowrite~\cite{dinesh2020retrowrite} & -- & Reassembleable disassembler & \xmark & \xmark & \xmark & \xmark \\
\stochfuzz~\cite{zhang2021stochfuzz} & -- & Trampoline & \cmark & \xmark & \xmark & \xmark \\
\shaderow \zafl~\cite{nagy2021breaking} & Zipr~\cite{hawkins2017zipr} & Direct rewriter & \cmark & \xmark & \xmark & \xmark \\
\midrule
\pear & \ddisasm~\cite{flores2020datalog} & Reassembleable disassembler & \cmark & \cmark & \cmark & \cmark \\
\bottomrule
\end{tabular}
\end{table*}

Without access to source code, a fuzzer must rely on static or dynamic binary
instrumentation for guidance.  While emulation frameworks (e.g.,
QEMU~\cite{bellard2005qemu}, \icicle~\cite{chesser2023icicle}) are commonly used
to fuzz targets compiled for different architectures, they suffer from
significant reductions in fuzzer throughput.

In contrast, \emph{static binary instrumentation} (SBI) does not suffer from the
same throughput reductions.  However, SBI is less accurate and reliable, because
recovering arbitrary control flow from a binary is undecidable.  Unfortunately,
this impacts the usability of SBI: \citet{shulte2022rewritercomp} compared ten
SBI frameworks and found that---despite broad support for \aflpp
instrumentation---only four successfully instrumented \emph{any} of the test
binaries.  \Citet{kim2023reassembly} undertook a similar study with comparable
results.  These studies highlight the ongoing challenges of SBI.

We summarize and compare five popular SBI fuzzing frameworks in
\cref{tab:sbi-compare}.
E9AFL, \afldyninst, and \stochfuzz rewrite target instructions
with a \emph{trampoline}, a small snippet of code that redirects control flow to
the instrumentation (and ensures control flow returns to the original code).
While trampolines are performant, they may not always be possible; e.g., if the
target instruction(s) is smaller than the trampoline.
\retrowrite uses \emph{reassembleable disassembly}, exploiting position
independent code~(PIC) and relocation data to generate an assembly file that can
be instrumented with fuzzer instrumentation.  However, not all targets are
compiled with PIC.
Finally,~\zafl lifts the binary into an intermediate representation~(IR), rewrites
this~IR, and then lowers it back down to native code.  While performant, the
rewriter may fail if the lifter encounters unsupported instructions.

\Cref{tab:sbi-compare} also shows the fuzzing features supported by these five
frameworks.  While four frameworks support the forkserver model,
none of them support the other features described in
\cref{sec:background-greybox-fuzzing}.  In contrast, \aflpp's QEMU mode
supports all of these features (albeit with larger performance overheads).  This
lack of modern fuzzer features---combined with the unreliability of existing SBI
frameworks---motivates us to develop \pear, which we describe in the following
section.

\section{\pear}
\label{sec:pear}

\begin{figure*}
\centering
\includegraphics[width=0.75\linewidth,keepaspectratio]{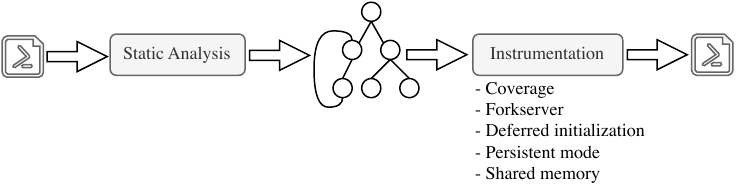}
\caption{\pear overview.  The target binary is disassembled and statically
analyzed before being instrumented and rewritten.}
\label{fig:pear-overview}
\end{figure*}

We present a high-level overview of \pear in \cref{fig:pear-overview}.  \pear
instruments target binaries using \ddisasm~\cite{flores2020datalog} and the
GrammaTech Intermediate Representation for Binaries~\cite{schulte2020gtirb}
(GTIRB).  First, \ddisasm disassembles the input binary and decodes a
superset of possible instructions to create a set of Datalog facts.  These facts
are then analyzed (e.g., for symbolization and to determine control flow) and
refined before being translated to GTIRB. 
We then use the \gtirbrewriting framework~\cite{gtirb-rewriting} to  
insert coverage instrumentation, the forkserver, and the code necessary to enable deferred initialization, persistent mode, and shared memory fuzzing.
The remainder of this section describes these steps in greater depth.

\subsection{Static Analysis}

\pear uses reassembleable disassembly~\cite{WangWW15} to achieve efficient static binary
instrumentation.  We use \ddisasm---a fast and accurate disassembler implemented
using Datalog---to disassemble the target binary and translate it to the GTIRB
intermediate representation for further analysis and instrumentation.  We choose
\ddisasm because of its high success rates in a range of dissasembly and
rewriting benchmarks~\cite{shulte2022rewritercomp,kim2023reassembly,%
  priyadarshan2023disassembly}.  Indeed, our evaluation reinforces these
findings, with \pear successfully instrumenting the vast majority of \fuzzbench
targets.  
The resulting GTIRB output (a serialized protocol buffer) contains all of the
information required for inserting fuzzing instrumentation.

\subsection{Instrumentation}

\pear's instrumentation consists of: coverage tracing, deferred initialization,
persistent mode, and shared memory fuzzing.  This instrumentation
supports~\verb|x64| Linux~ELF and Windows~PE executables and is compatible with
the state-of-the-art \aflpp fuzzer.  We describe the design and implementation
of this instrumentation in the following sections.

\subsubsection{Coverage Tracing}
\label{sec:coverage-impl}

\pear tracks edge coverage by inserting trampolines to a tracer function at
the start of a GTIRB basic block.\footnote{Unfortunately, \gtirbrewriting does
  not support inserting instrumentation code using non-live registers at patch
  locations, which is required for sound insertion of inlined instrumentation.}
This approach is inspired by the \verb|__afl_maybe_log| function from AFL++'s
assembly-level instrumentation.  Like \verb|__afl_maybe_log|, \pear's tracing
function takes a single argument: a random integer identifying the current
block.  \Cref{lst:trampoline} shows our trampoline code, where \verb|<BLOCK_ID>|
(\cref{line:block-id}) is the basic block identifier.

\begin{minipage}{\linewidth}
\begin{lstlisting}[style=x86assembly, caption={Basic block trampoline.},
  label={lst:trampoline}]
lea    rsp,[rsp-0x98]
mov    QWORD PTR [rsp], rdx
mov    QWORD PTR [rsp+0x8], rcx
mov    QWORD PTR [rsp+0x10], rax
mov    rcx, <BLOCK_ID> (^ \label{line:block-id} ^)
call   __afl_trace
mov    rax, QWORD PTR [rsp+0x10]
mov    rcx, QWORD PTR [rsp+0x8]
mov    rdx, QWORD PTR [rsp]
lea    rsp, [rsp+0x98]
\end{lstlisting}
\end{minipage}

Unlike AFL++---which represents the
target as a collection of \emph{intraprocedural} control-flow graphs~(CFG)---GTIRB
represents the target as a single \emph{interprocedural}~CFG, where blocks can
be terminated by call instructions (not just branches) and ``fallthrough''
edges connect the caller block to a successor block (containing the instructions
following the call).  To avoid unnecessary instrumentation, \pear does not
instrument blocks with an incoming fallthrough edge.

\Cref{lst:edge-trace} shows the \verb|__afl_trace| function called by the
trampoline.  This function follows the ``classic'' edge coverage approach used
by AFL, xor-ing the previous block identifier (\verb|__afl_prev_loc|) with the
current block identifier (passed in~\verb|rcx|).  The result of this xor---an
edge identifier---is used as a lookup into the coverage map (pointed to by
\verb|__afl_area_ptr|) where an edge counter is incremented.

\begin{minipage}{\linewidth}
\begin{lstlisting}[style=x86assembly, caption={Edge coverage tracing.},
label={lst:edge-trace}]
; Save state
lahf
seto    al

; Record path in coverage map
mov    rdx,[rip + __afl_area_ptr]
xor    rcx, QWORD PTR [rip + __afl_prev_loc]
xor    QWORD PTR [rip + __afl_prev_loc], rcx
shr    QWORD PTR [rip + __afl_prev_loc], 1
inc    BYTE PTR [rdx + rcx * 1]
adc    BYTE PTR [rdx + rcx * 1], 0x0

; Restore state
add    al,0x7f
sahf

ret
\end{lstlisting}
\end{minipage}

\subsubsection{Initialization}

\paragraph*{Coverage Map Setup.}
\Cref{lst:__afl_setup} shows the initialization routine inserted by
\pear.\footnote{This code is written in~C, compiled into its own object file,
and later linked with the instrumented target.}  A call to this routine is
inserted at the target's entrypoint.  This routine sets up a dummy map area
(allowing the target to run without a fuzzer) before attaching to the shared
memory map (\cref{line:attach}). Per \cref{sec:coverage-impl}, this shared memory
map stores edge hit counts and is used to communicate coverage information
to the fuzzer.  Notably, \gtirbrewriting does not support adding data to the
target's \verb|.bss| section, forcing \pear to use a dynamic memory allocation
for the dummy map area (\cref{line:dummy-map-alloc}).

\begin{minipage}{\linewidth}
\begin{lstlisting}[language=C,
  caption={Coverage map initialization.},
  label={lst:__afl_setup}]
void __afl_setup(void) {
  __afl_area_ptr_dummy =  malloc(0x10000);(^ \label{line:dummy-map-alloc} ^)
  if (__afl_area_ptr_dummy == NULL) {
    perror("AFL dummy map allocation failed");
    _exit(1);
  }

  if (getenv("AFL_DEBUG"))
    __afl_debug = 1;

  __afl_map_shm();(^ \label{line:attach} ^)
}
\end{lstlisting}
\end{minipage}

\paragraph*{Deferred Initialization.}
\aflpp recognizes a deferred forkserver by looking for the
\verb|#SIG_AFL_DEFER_FORKSRV#| string in the target binary.  \pear allows the
user to specify the forkserver's location (as a function or address).  This
location defines when the target process has been fully initialized and can be
cloned.  The forkserver ensures the target is only initialized once, rather than
during the execution of every input.  Notably, \pear does not provide a
forkserver on Windows due to the lack of support for \texttt{fork}
operations~\cite{WinnieFuzzer}.

\subsubsection{Persistent Mode}
\label{sec:pear-persistent-mode}

\begin{figure}
\centering

\begin{subfigure}{\linewidth}
\includegraphics[width=\linewidth,keepaspectratio]{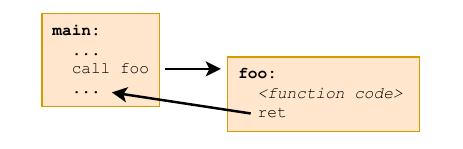}
\caption{Original function call.}
\label{fig:pear_persistent_mode_before}
\end{subfigure}

\begin{subfigure}{\linewidth}
\includegraphics[width=\linewidth,keepaspectratio]{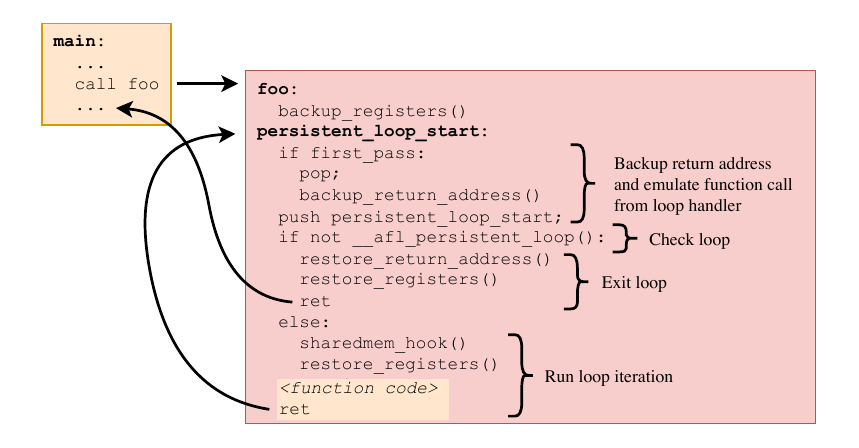}
\caption{After persistent mode application.}
\label{fig:pear_persistent_mode_after}
\end{subfigure}

\caption{Example of \pear's persistent mode instrumentation.}
\label{fig:pear_persistent_mode}
\end{figure}

\Cref{fig:pear_persistent_mode} summarizes \pear's persistent mode
instrumentation, while \cref{lst:persistent_patch} shows the corresponding
assembly code.  In \cref{fig:pear_persistent_mode_before}, the function
\texttt{foo} is selected as a fuzz target.  If source code were available, we
would wrap the call to \texttt{foo} within a loop---e.g., using \aflpp's
\verb|__afl_persistent_loop| macro---signaling to the fuzzer that~\texttt{foo}
should be executed with multiple inputs \emph{without} spawning a new proess per
input.

\Cref{lst:persistent_patch} shows the persistent mode handler inserted
at~\texttt{foo}'s entrypoint.  The handler first saves the current
register state before entering the persistent loop.  The first time the loop is
entered, the caller's return address (in \texttt{main}) is saved to memory
(\crefrange{line:first-pass-start}{line:first-pass-end}).  In the loop body,
the start address of the loop replaces the caller's return address
(\crefrange{line:not-first-pass-start}{line:not-first-pass-end}), ensuring the
persistent loop is always executed.  After looping for~\verb|<PERSISTENT_MODE_COUNT>|
iterations (\crefrange{line:check-loop-start}{line:check-loop-end}) the original
return address and register state are restored
(\crefrange{line:restore-state-start}{line:restore-state-end}).

\begin{lstlisting}[style=x86assembly, caption={Persistent mode handler.},
label={lst:persistent_patch}]
; Backup register state (omitted for brevity)

; Start of persistent loop
.Lsetup_loop:
movzx    eax, BYTE PTR first_pass[rip]
test     al, al
je       .Lnot_first_pass

; On first pass, save and overwrite the return address
; (first_pass is set in __afl_persistent_loop)
pop     rax(^ \label{line:first-pass-start} ^)
mov     QWORD PTR [rip + p_mode_ret_addr_backup], rax(^ \label{line:first-pass-end} ^)

.Lnot_first_pass:
; On subsequent passes, set new return address
lea     rax, [rip + .Lsetup_loop](^ \label{line:not-first-pass-start} ^)
push    rax(^ \label{line:not-first-pass-end} ^)

; Check whether to continue loop
mov     rcx, rsp(^ \label{line:check-loop-start} ^)
lea     rsp, [rsp - 0x80]
and     rsp, 0xfffffffffffffff0
push    rcx
push    rcx
mov     edi, <PERSISTENT_MODE_COUNT>
call    __afl_persistent_loop
pop     rcx
mov     rsp, rcx

test    eax,eax
jne     .Lstart_func(^ \label{line:check-loop-end} ^)

; Break from loop, restoring the original return address
mov     rax, QWORD PTR [rip + p_mode_ret_addr_backup](^ \label{line:restore-state-start} ^)
lea     rsp, [rsp + 0x8]
push    rax
; Restore register state (omitted for brevity)
ret(^ \label{line:restore-state-end} ^)

; The function to fuzz
.Lstart_func:
\end{lstlisting}


\subsubsection{Shared Memory Fuzzing}
\label{sec:pear-shared-mem-fuzz}
\label{sec:prop-fuzzing}

\pear implements shared memory fuzzing using a shared memory hook, similar to
\aflpp's QEMU mode.  \Cref{lst:pear_shmem_hook} shows the hook's function
signature, which takes the following arguments:
\begin{inlineroman}
\item the current state of program registers;
\item a pointer to the current input; and
\item the length of the input.
\end{inlineroman}

\begin{minipage}{\linewidth}
\begin{lstlisting}[language=C, caption={Shared memory hook.}, label={lst:pear_shmem_hook}]
struct __attribute__((__packed__)) x86_64_regs {
  uint64_t rax, rbx, rcx, rdx, rdi, rsi, 
      r8, r9, r10, r11, r12, r13, r14, r15;
  uint8_t xmm_regs[16][16];
};

void __afl_rewrite_sharedmem_hook(
  struct x86_64_regs *saved_regs,
  uint8_t *input_buf,
  uint32_t input_buf_len
);
\end{lstlisting}
\end{minipage}

The hook's implementation depends on how the target processes the input and
where in the program the hook function is called.  When instrumenting a target
the user can select one of the following locations for the hook call:
\begin{itemize}
\item At the beginning of the persistent mode loop;
\item At a user-selected location, specified through code address or function name; or
\item Immediately after forkserver initialization.
\end{itemize}

As \pear operates at a binary level, there is no source-level mechanism (such
as AFL++'s \texttt{\_\_AFL\_FUZZ\_TESTCASE\_BUF} macro) to redirect how the
target reads its input.

Instead, \pear saves the target's register state to a global data section
(\verb|p_mode_reg_backup|) before invoking the hook, and restores registers
from here afterwards. This allows the hook to modify register values through
the \verb|x86_64_regs| struct - for example, redirecting a function's pointer
argument to the shared memory test case. The modified register value then takes
effect when the target function executes.
\cref{lst:shmem_hook} shows two example hooks. In both cases, the target
function takes a buffer pointer as its first argument (\verb|rdi| under the
System V ABI). The first hook redirects \verb|rdi| to point directly at AFL++'s
shared memory test case, avoiding any copy. The second hook instead copies the
test case into the target's existing buffer, which is necessary if the target
expects the test case to be in a specific area of memory. The hook is compiled
as an object file and statically linked with the instrumented binary.

\begin{minipage}{\linewidth}
\begin{lstlisting}[language=C, caption={Example shared memory hooks.}, label={lst:shmem_hook}]

// Hook for LLVMFuzzerTestOneInput(uint8_t *Data, size_t Size), where we can redirect rdi to point at shared memory test case.
void __pear_sharedmem_hook(struct x86_64_regs *regs,
    uint8_t *input_buf, uint32_t input_buf_len) {
  regs->rdi = (uint64_t) input_buf;
  regs->rsi = input_buf_len;
}

// Hook for process(char* buf), where buffer points to a caller-owned 100-byte buffer that cannot be redirected.
void __pear_sharedmem_hook(struct x86_64_regs *regs,
    uint8_t *input_buf, uint32_t input_buf_len) {
  memset((void *)regs->rdi, 0, 100);
  if (input_buf_len > 100) input_buf_len = 100;
  memcpy((void *)regs->rdi, input_buf, input_buf_len);
}
\end{lstlisting}
\end{minipage}

\section{Evaluation}
\label{sec:evaluation}

We evaluate \pear to determine \emph{the feasibility of implementing modern fuzzing
techniques using SBI frameworks without compromising on performance}.
We conduct our evaluation on the \fuzzbench~\cite{metzman2021fuzzbench}
benchmark suite, comparing \pear against other
state-of-the-art binary-only fuzzers. 

Our evaluation focuses on the following metrics:
\begin{itemize}[leftmargin=*]
\item \textbf{Robustness}: the ability of a binary-only fuzzer to successfully
instrument a given target.

\item \textbf{Code coverage}: the amount of code covered by the fuzzer on a
particular target.

\item \textbf{Speed}: the number of iterations achieved (per unit of time) on a
particular target.

\end{itemize}
We evaluate these metrics in \fuzzbench
(\cref{sec:exp-robustness,sec:exp-coverage,sec:exp-ablation}).
Our evaluation focuses on \pear's Linux
ELF instrumentation and does not include \pear's Windows~PE support (due to the
lack of support for~PE files in \fuzzbench).

\subsection{Experiment 1: Robustness}
\label{sec:exp-robustness}

SBI-based fuzzers have traditionally failed to instrument complex ``real-world''
targets.  Thus, we compare \pear's ability to instrument all~25 targets in the
\fuzzbench benchmark suite, comparing \pear to \numevalsbi other state-of-the-art
SBI-based fuzzers: \zafl~\cite{nagy2021breaking},
\stochfuzz~\cite{zhang2021stochfuzz}, \afldyninst~\cite{vanheuser2021afldyninst},
and E9AFL~\cite{gao2021scalable}.  We exclude
\retrowrite~\cite{dinesh2020retrowrite} because it only supports
position-independent executables~(PIE), which \fuzzbench does not produce.

We use \afldyninst's ``experimental performance mode'' as recommended by the
authors, which led to more targets being instrumented (compared to \afldyninst's
default mode).  For two targets (\texttt{curl} and \texttt{libjpeg-turbo}),
\pear required manual \ddisasm hints to produce correct instrumentation:
\texttt{curl} contained data tables in its \texttt{.text} section that were
misidentified as code, while \texttt{libjpeg-turbo} had a constant
misidentified as a symbolic expression.

\Cref{tab:succ_bench} summarizes the \fuzzbench targets each fuzzer was able to
instrument.  Failing builds fell into one of two categories:
\begin{itemize}[leftmargin=*]
\item \textbf{Failed application}. The tool failed to produce an instrumented
binary (e.g., due to a build or rewriting failure).

\item \textbf{Faulty instrumentation}. The tool produced an instrumented target.
However, these binaries crashed on benign inputs, indicating an error in the
instrumentation.
\end{itemize}

\begin{table}
\centering
\footnotesize
\caption{Successfully instrumented \fuzzbench targets.
\cmark~indicates successful instrumentation,
\cmark\textsuperscript{*}~indicates success with manual \ddisasm hints,
\xmark~indicates faulty instrumentation, and
\xmark\xmark~indicates no instrumented binary was generated.
}
\label{tab:succ_bench}

\begin{adjustbox}{width=\linewidth}
\begin{tabular}{lccccc}
\toprule
Target                                       & \pear         & \zafl         & \stochfuzz    & \afldyninst   & E9AFL         \\
\midrule
\shaderow \texttt{bloaty\_fuzz\_target}                & \xmark        & \cmark        & \cmark        & \xmark        & \xmark        \\
\texttt{curl\_curl\_fuzzer\_http}            & \cmark\textsuperscript{*} & \cmark & \cmark & \cmark        & \xmark        \\
\shaderow \texttt{freetype2\_ftfuzzer}                 & \cmark        & \cmark        & \cmark        & \cmark        & \xmark        \\
\texttt{harfbuzz\_hb-shape-fuzzer}           & \cmark        & \cmark        & \cmark        & \cmark        & \xmark        \\
\shaderow \texttt{jsoncpp\_jsoncpp\_fuzzer}            & \cmark        & \cmark        & \cmark        & \xmark        & \xmark        \\
\texttt{lcms\_cms\_transform\_fuzzer}        & \cmark        & \cmark        & \xmark        & \cmark        & \cmark        \\
\shaderow \texttt{libjpeg-turbo\_libjpeg\_turbo\_fuzzer} & \cmark\textsuperscript{*} & \cmark & \cmark & \cmark & \xmark \\
\texttt{libpcap\_fuzz\_both}                 & \cmark        & \cmark        & \cmark        & \cmark        & \xmark        \\
\shaderow \texttt{libpng\_libpng\_read\_fuzzer}        & \cmark        & \cmark        & \cmark        & \cmark        & \xmark        \\
\texttt{libxml2\_xml}                        & \cmark        & \cmark        & \cmark        & \cmark        & \xmark        \\
\shaderow \texttt{libxslt\_xpath}                      & \cmark        & \xmark \xmark & \cmark        & \cmark        & \xmark        \\
\texttt{mbedtls\_fuzz\_dtlsclient}           & \cmark        & \cmark        & \cmark        & \cmark        & \xmark        \\
\shaderow \texttt{mruby\_mruby\_fuzzer\_8c8bbd}       & \cmark        & \cmark        & \cmark        & \xmark        & \xmark        \\
\texttt{openh264\_decoder\_fuzzer}           & \cmark        & \cmark        & \cmark        & \cmark        & \xmark        \\
\shaderow \texttt{openssl\_x509}                       & \cmark        & \cmark        & \cmark        & \cmark        & \xmark        \\
\texttt{openthread\_ot-ip6-send-fuzzer}      & \cmark        & \cmark        & \cmark        & \xmark        & \xmark        \\
\shaderow \texttt{php\_php-fuzz-parser\_0dbedb}        & \cmark        & \xmark \xmark & \xmark \xmark & \xmark        & \xmark        \\
\texttt{proj4\_proj\_crs\_to\_crs\_fuzzer}   & \xmark        & \cmark        & \xmark \xmark & \xmark        & \xmark        \\
\shaderow \texttt{re2\_fuzzer}                         & \cmark        & \cmark        & \cmark        & \xmark        & \xmark        \\
\texttt{sqlite3\_ossfuzz}                    & \xmark        & \cmark        & \cmark        & \xmark        & \xmark        \\
\shaderow \texttt{stb\_stbi\_read\_fuzzer}             & \cmark        & \cmark        & \cmark        & \xmark        & \xmark        \\
\texttt{systemd\_fuzz-link-parser}           & \cmark        & \cmark        & \cmark        & \cmark        & \xmark        \\
\shaderow \texttt{vorbis\_decode\_fuzzer}              & \cmark        & \cmark        & \cmark        & \cmark        & \cmark        \\
\texttt{woff2\_convert\_woff2ttf\_fuzzer}    & \cmark        & \cmark        & \cmark        & \xmark        & \xmark        \\
\shaderow \texttt{zlib\_zlib\_uncompress\_fuzzer}      & \cmark        & \cmark        & \cmark        & \cmark        & \xmark        \\
\cmidrule(lr){1-6}
\# Successful Instrumentations               & 22            & 23            & 22            & 15            & 2             \\
\bottomrule
\end{tabular}
\end{adjustbox}
\end{table}

\begin{hlbox}{Result 1}
\zafl, \pear, and \stochfuzz are the most robust SBI-based fuzzers, instrumenting
23, 22, and 22 out of 25 \fuzzbench targets respectively. However, \pear is the
only SBI fuzzer that supports deferred initialization, persistent mode, and
shared memory fuzzing (\cref{tab:sbi-compare}).
\end{hlbox}

\subsection{Experiment 2: Coverage and Performance}
\label{sec:exp-coverage}

Here we evaluate the code coverage and throughput achieved by \pear, comparing
\pear against other state-of-the-art binary-only fuzzers on a large set of
\fuzzbench targets.

\paragraph*{Fuzzer Selection.}

We compare five SBI-based fuzzers: \pear (base configuration),
\zafl~\cite{nagy2021breaking}, \stochfuzz~\cite{zhang2021stochfuzz},
\afldyninst~\cite{vanheuser2021afldyninst}, and E9AFL~\cite{gao2021scalable}.
As a DBI baseline, we include \aflpp's QEMU mode with persistent mode and
shared memory enabled, representing the strongest available DBI configuration.
As a compiler baseline, we include \aflpp with compiler instrumentation (using
\fuzzbench 's \texttt{aflplusplus} configuration), which serves as a performance upper bound.  \pear's
persistent mode and shared memory variants are evaluated separately in the
ablation study (\cref{sec:exp-ablation}).

\paragraph*{Experimental Setup.}

We selected 20~\fuzzbench\footnote{We forked \fuzzbench at
commit \texttt{2a2ca6ae4c5d171a52b3e20d9b7a72da306fe5b8} to incorporate our
fuzzers.} targets: all targets from the benchmark suite for which at least one
binary-only fuzzer could produce a working instrumented binary. 

We excluded the targets that \pear failed to instrument (\texttt{bloaty},
\texttt{proj4}, \texttt{sqlite3}). We also excluded the bug-finding benchmarks
(\texttt{php} and \texttt{mruby}) as \fuzzbench does not support running bug
and coverage benchmarks in the same experiment.
Each fuzzer-benchmark pair was fuzzed for~\qty{24}{\hour} across ten trials.
This experiment was run on a local server with four Intel Xeon Gold~6252
\qty{2.10}{\giga\hertz}~CPUs, \qty{187}{\gibi\byte} of~RAM, and running
Ubuntu~20.04~LTS.
The total compute for this experiment was approximately~\totalfuzztime.

\subsubsection{Results}

\Cref{tab:coverage} presents the per-benchmark median final edge coverage,
normalized to \aflpp with compiler instrumentation.  Among the SBI-based
fuzzers, \pear achieves the highest average normalized coverage
score~(\qty{87.91}{\percent}), comparable to \aflpp QEMU
mode~(\qty{87.84}{\percent}).  \zafl follows at~\qty{85.17}{\percent},
then \stochfuzz~(\qty{79.20}{\percent}),
\afldyninst~(\qty{57.79}{\percent}), and E9AFL~(\qty{6.62}{\percent}).
\pear with persistent mode and shared memory fuzzing achieves the best or
tied-best coverage among SBI fuzzers on the majority of benchmarks,
demonstrating that modern SBI frameworks can match DBI approaches in coverage
effectiveness.

\begin{figure*}
\centering
\includegraphics[width=\linewidth]{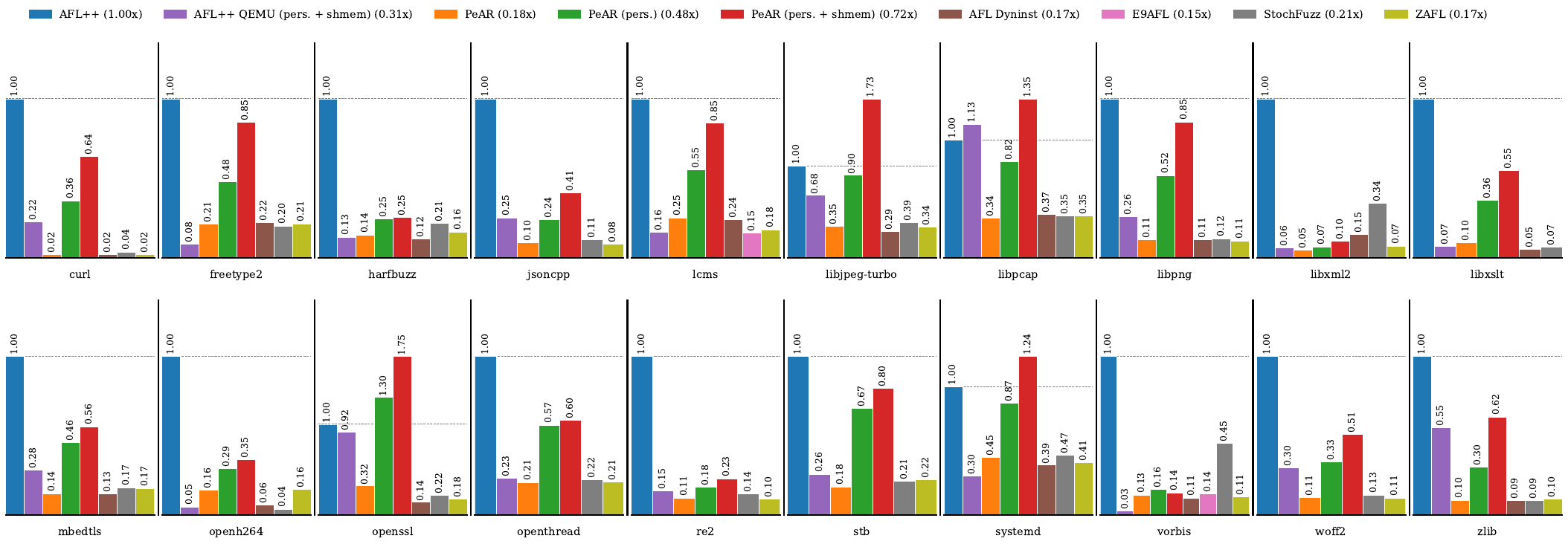}
\caption{Throughput of each fuzzer per benchmark, expressed as a ratio to
\aflpp with compiler instrumentation.  Higher is better; the dashed line
marks parity with \aflpp.}
\label{fig:throughput}
\end{figure*}

\Cref{fig:throughput} shows the throughput of each fuzzer on every benchmark,
normalized to \aflpp with compiler instrumentation.  Among the binary-only
fuzzers, \pear with persistent mode and shared memory fuzzing consistently
achieves the highest throughput ratio, while base \pear remains competitive
with \zafl and \stochfuzz.

\begin{table}[t]
\centering
\caption{Median final edge coverage per benchmark, normalized to AFL++ with compiler instrumentation. \textbf{Bold} indicates the best SBI-based fuzzer per benchmark. ``---'' indicates the fuzzer could not instrument the target.}
\label{tab:coverage}
\tiny
\setlength{\tabcolsep}{2.5pt}
\begin{tabular}{lrrrrrrrrr}
\toprule
Benchmark & \rotatebox{0}{\texttt{AFL++}} & \rotatebox{0}{\texttt{AFL++ QEMU}} & \rotatebox{0}{\texttt{PeAR}} & \rotatebox{0}{\texttt{PeAR p.}} & \rotatebox{0}{\texttt{PeAR p.+s.}} & \rotatebox{0}{\texttt{AFL Dyninst}} & \rotatebox{0}{\texttt{E9AFL}} & \rotatebox{0}{\texttt{StochFuzz}} & \rotatebox{0}{\texttt{ZAFL}} \\
\midrule
\texttt{curl} & 1.00 & 0.96 & 0.93 & 0.96 & \textbf{0.98} & 0.84 & --- & 0.77 & 0.94 \\
\texttt{freetype2} & 1.00 & 0.67 & 0.71 & 0.72 & \textbf{0.73} & 0.34 & --- & 0.58 & 0.71 \\
\texttt{harfbuzz} & 1.00 & 0.96 & 0.96 & 0.97 & \textbf{0.98} & 0.95 & --- & 0.95 & 0.96 \\
\texttt{jsoncpp} & 1.00 & 1.00 & \textbf{1.00} & \textbf{1.00} & \textbf{1.00} & --- & --- & \textbf{1.00} & \textbf{1.00} \\
\texttt{lcms} & 1.00 & 0.33 & 0.33 & 0.60 & 0.53 & 0.31 & 0.33 & --- & \textbf{0.74} \\
\texttt{libjpeg-turbo} & 1.00 & 1.00 & \textbf{1.00} & \textbf{1.00} & \textbf{1.00} & \textbf{1.00} & --- & \textbf{1.00} & \textbf{1.00} \\
\texttt{libpcap} & 1.00 & 0.01 & \textbf{0.01} & \textbf{0.01} & \textbf{0.01} & \textbf{0.01} & --- & \textbf{0.01} & \textbf{0.01} \\
\texttt{libpng} & 1.00 & 1.00 & \textbf{1.00} & \textbf{1.00} & \textbf{1.00} & 0.99 & --- & 0.93 & \textbf{1.00} \\
\texttt{libxml2} & 1.00 & 0.96 & 0.96 & 0.96 & \textbf{0.97} & 0.82 & --- & 0.96 & 0.96 \\
\texttt{libxslt} & 1.00 & 0.97 & 0.98 & 0.98 & \textbf{0.99} & 0.64 & --- & 0.76 & --- \\
\texttt{mbedtls} & 1.00 & 0.98 & 0.98 & \textbf{1.00} & \textbf{1.00} & 0.94 & --- & 0.75 & 0.99 \\
\texttt{openh264} & 1.00 & 0.98 & 0.98 & 0.99 & \textbf{1.00} & 0.96 & --- & 0.81 & 0.99 \\
\texttt{openssl} & 1.00 & 1.00 & \textbf{1.00} & \textbf{1.00} & \textbf{1.00} & \textbf{1.00} & --- & \textbf{1.00} & \textbf{1.00} \\
\texttt{openthread} & 1.00 & 0.94 & 0.95 & 0.96 & \textbf{1.00} & --- & --- & 0.92 & 0.93 \\
\texttt{re2} & 1.00 & 0.99 & 0.99 & \textbf{1.00} & \textbf{1.00} & --- & --- & 0.99 & \textbf{1.00} \\
\texttt{stb} & 1.00 & 0.95 & \textbf{0.95} & \textbf{0.95} & \textbf{0.95} & --- & --- & 0.93 & \textbf{0.95} \\
\texttt{systemd} & 1.00 & 0.92 & 0.92 & 0.92 & 0.92 & 0.88 & --- & \textbf{0.93} & 0.92 \\
\texttt{vorbis} & 1.00 & 0.99 & 0.98 & 0.99 & 0.99 & 0.97 & 0.99 & 0.65 & \textbf{1.00} \\
\texttt{woff2} & 1.00 & 0.96 & 0.95 & \textbf{0.96} & \textbf{0.96} & --- & --- & 0.92 & 0.94 \\
\texttt{zlib} & 1.00 & 1.00 & \textbf{1.00} & \textbf{1.00} & \textbf{1.00} & 0.94 & --- & 0.99 & 0.99 \\
\midrule
\textit{Average} & 1.00 & 0.88 & 0.88 & \textbf{0.90} & \textbf{0.90} & 0.77 & 0.66 & 0.83 & \textbf{0.90} \\
\bottomrule
\end{tabular}
\end{table}

\begin{hlbox}{Result 2}
\pear achieves the highest code coverage among SBI-based fuzzers, with a
normalized score of~\qty{87.91}{\percent}, comparable to \aflpp QEMU
mode~(\qty{87.84}{\percent}), while also achieving competitive throughput.
\end{hlbox}

\subsection{Experiment 3: Ablation Study}
\label{sec:exp-ablation}

A key advantage of \pear as a framework is its ability to implement advanced
fuzzing techniques---persistent mode and shared memory fuzzing---that are
unsupported by other SBI-based fuzzers (\cref{tab:sbi-compare}).  To isolate
the contribution of each feature, we compare three \pear configurations across
all~20 benchmarks:
\begin{itemize}[leftmargin=*]
\item \textbf{\pear} (base): forkserver with coverage tracing only.
\item \textbf{\pear persistent}: adds persistent mode, avoiding \texttt{fork}
overhead for each test case.
\item \textbf{\pear persistent + shmem}: adds shared memory fuzzing on top of
persistent mode, avoiding file~I/O for test case delivery.
\end{itemize}

\subsubsection{Results}

Persistent mode improves \pear's throughput by a median of~$2.56\times$ across
all~20 benchmarks (range: $1.28$--$17.57\times$).  Adding shared memory fuzzing
provides further improvement, achieving a median of~$4.07\times$ over the base
configuration (range: $1.12$--$31.23\times$).  The magnitude of improvement is
benchmark-dependent: targets with expensive initialization (e.g.,
\texttt{curl}: $17.57\times$ with persistent mode, $31.23\times$ with shared
memory) benefit the most, while targets with lightweight initialization (e.g.,
\texttt{vorbis}: $1.28\times$) see smaller gains.

These throughput improvements translate into coverage gains.  The average
normalized coverage score increases from~\qty{87.91}{\percent} (base)
to~\qty{89.84}{\percent} (persistent) to~\qty{89.95}{\percent} (persistent +
shared memory), demonstrating that the additional throughput allows \pear to
explore more of the target's state space within the 24-hour campaign.

\begin{hlbox}{Result 3}
Persistent mode improves \pear's throughput by a median of~$2.56\times$
across~20 benchmarks.  Adding shared memory fuzzing provides a further
improvement to a median of~$4.07\times$, demonstrating the value of \pear's
extensible framework architecture.
\end{hlbox}

\subsection{Summary of Findings}

We summarize here the results of our experiments along the three metrics listed at the beginning of this section.

\noindent \textbf{Robustness.}
\zafl, \pear, and \stochfuzz are the most robust SBI fuzzers, instrumenting 23,
22, and 22 out of 25 \fuzzbench targets respectively.  Unlike other SBI fuzzers,
\pear uniquely supports deferred initialization, persistent mode, and shared
memory fuzzing.

\noindent \textbf{Code coverage.}
\pear achieves the highest normalized coverage score among SBI-based fuzzers
(\qty{87.91}{\percent} to \qty{89.95}{\percent} depending on configuration),
comparable to \aflpp QEMU mode~(\qty{87.84}{\percent}).

\noindent \textbf{Speed.}
\pear with persistent mode and shared memory fuzzing achieves a median
throughput improvement of~$4.07\times$ over base \pear, and a median
of~$3.56\times$ over the next-fastest SBI fuzzer on each benchmark.
These throughput gains translate into measurable coverage improvements across
all 20~benchmarks.

\section{Conclusions and Future Work}
\label{sec:conclusions}

In light of recent advances in SBI frameworks such as GTIRB, we re-examine the
viability of using SBI to build an effective and extensible binary-only fuzzing
framework.  We show that despite the many challenges in implementing SBI fuzzing
frameworks~\cite{shulte2022rewritercomp,kim2023reassembly}, it is indeed
possible to build an SBI fuzzing framework that is robust, achieves comparable
code coverage to compiler instrumentation, and supports advanced fuzzing
techniques.

Our implementation, \pear, instruments 22 out of 25 \fuzzbench targets,
comparable to \zafl~(23/25) and \stochfuzz~(22/25).  However, unlike these
tools, \pear is the only SBI framework that supports deferred initialization,
persistent mode, and shared memory fuzzing.  Our ablation study demonstrates
the value of this extensibility: persistent mode and shared memory fuzzing
provide a median throughput improvement of~$4.07\times$, with improvements up
to~$31\times$ on targets with expensive initialization.

For future work, we plan to implement support for a wider range of architetures and platforms. An experimental version of \pear for Windows PE binary is currently in development. We also plan to add other advanced instrumentations, in particular, adding support to instrument sanitizers into binaries, similar to \retrowrite.

\bibliographystyle{ACM-Reference-Format}
\bibliography{bib}

\end{document}